\documentclass[aps,pra,showpacs,twocolumn,amsmath,amssymb]{revtex4}
\usepackage{graphicx,bbm}

\newcommand{\un}[1]{{\underline{#1}}}
\newcommand{\bra}[1]{{\langle #1 \vert}}

\newcommand{\ket}[1]{{\vert #1 \rangle}}
\newcommand{\xket}[1]{{\vert #1 \rangle}}
\newcommand{\braket}[2]{\langle #1 \vert #2 \rangle}

\newcommand{\ii}{ {\rm i} }
\newcommand{\dd}{ {\rm d} }
\newcommand{\ZZ}{\mathbb{Z}}

\newcommand{\CC}{\mathbb{C}}
\newcommand{\cc}{ {\hat c} }
\newcommand{\bb}{ {\hat b} }
\newcommand{\aaa}{ {\hat a} }

\newcommand{\NESS}{{\rm ness}}
\newcommand{\LL}{{\hat {\cal L}}}

\newcommand{\mm}[1]{{\mathbf{#1}}}
\newcommand{\half}{\frac{1}{2}}

\def\tr{{\,{\rm tr}\,}}

\def\one{\mathbbm{1}}
\def\re{{\,{\rm Re}\,}}

\def\ness{\xket{\NESS}}

\begin{document}

\title{Quantum phase transition in a far from equilibrium steady 
state of XY spin chain}
\author{Toma\v{z} Prosen and Iztok Pi\v{z}orn}
\affiliation{Department of physics, FMF, University of Ljubljana,
Jadranska 19, SI-1000 Ljubljana, Slovenia}


\begin{abstract}
Using quantization in the Fock space of operators we compute the
non-equilibrium steady state in an open Heisenberg XY spin 1/2 chain of finite 
but large size coupled to Markovian baths at its ends. Numerical and
theoretical evidence is given for a far from equilibrium quantum phase 
transition with spontaneous emergence of long-range order in 
spin-spin correlation functions, characterized by a transition from 
saturation to linear growth with the size of the entanglement entropy in operator space.\!
\end{abstract}


\maketitle

Non-perturbative physics of {\em many-body} {\em open} quantum systems 
{\em far from equilibrium} is largely an unexplored field. 
In one-dimensional locally interacting quantum systems equilibrium phase 
transitions -- {\em quantum phase transitions} (QPT) -- 
can occur at zero temperature only and are by now well understood 
\cite{sachdev}. QPT are typically characterized by vanishing of the Hamiltonian's spectral gap 
in the thermodynamic limit at the critical point, and
(logarithmic) enhancement of the entanglement entropy and other measures of
quantum correlations in the ground state \cite{amico}.
Much less is known about the physics of QPT out of equilibrium,
studies of which have been 
usually limited to near equilibrium regimes or using involved and 
approximate analytical techniques (e.g. \cite{feldman,takei_kim}). 

There exist two general theoretical approaches to a description
of non-equilibrium open quantum systems, namely the non-equilibrium 
Green's function method \cite{kamenev}, and the
quantum master equation \cite{lindblad,breuer}.
In this Letter we adpot the latter and present a quasi-exactly solvable example of an open 
Heisenberg XY spin $1/2$ chain exhibiting a novel type of phase transition far
from equilibrium; characterized by a sudden 
appearance of long-range magnetic order in {\em non-equilibrium steady state} 
(NESS) as the magnetic field is reduced, and the transition from saturation to 
linear growth with size of the {\em operator space entanglement entropy} 
(OSEE) of NESS.

The Hamiltonian of the quantum XY chain reads
\begin{equation}
H = \sum_{m=1}^{n-1} \left( \frac{1+\gamma}{2} 
\sigma^x_m \sigma^x_{m+1} + \frac{1-\gamma}{2} \sigma^y_m \sigma^y_{m+1}\right)
+ \sum_{m=1}^n h \sigma^z_m
\label{eq:hamXY}
\end{equation}
where $\sigma^{x,y,z}_m,m=1,\ldots,n$ are Pauli operators acting on a 
string of $n$ spins. We may assume that parameters $\gamma$ (anisotropy) and $h$ (magnetic field) are
{\em non-negative}. It is known that XY model (\ref{eq:hamXY}) exhibits
(equilibrium) {\em critical} behavior in the thermodynamic limit $n\to\infty$ along the 
lines: $\gamma=0, h \le 1$, and $h=1$.
Here we consider an {\em open} XY chain whose density matrix evolution $\rho(t)$ is
governed by the Lindblad master equation \cite{lindblad} (we set $\hbar=1$)
\begin{equation}
\frac{\dd\rho }{\dd t} = \LL \rho :=
-\ii [H,\rho] + \sum_{\mu=1}^M \left(2 L_\mu \rho L_\mu^\dagger - \{L_\mu^\dagger L_\mu,\rho\} \right)
\label{eq:lind}
\end{equation}
and study a phase transition in NESS. The simplest nontrivial bath (Lindblad) operators 
acting only on the first and the last spin are chosen ($M=4$)
\begin{equation}
L_{1,2} = \sqrt{\Gamma_{1,2}^{\rm L}} \sigma^{\mp}_1, \qquad
L_{3,4} = \sqrt{\Gamma_{1,2}^{\rm R}} \sigma^{\mp}_n,  
\label{eq:bathsc}
\end{equation}
where $\sigma^\pm_m=(\sigma^x_m \pm \ii \sigma^y_m)/2$ \cite{wichterich}.
For $h\gg 1$, the ratios $\Gamma^\lambda_2/\Gamma^\lambda_1 = \exp(-2h/T_{\lambda})$ are simply related to canonical
temperatures of the end spins $T_\lambda$, $\lambda={\rm L},{\rm R}$.

Note that Lindblad equation (\ref{eq:lind}) can be rigorously derived within the so-called 
Markov approximation \cite{breuer}
which is justified for macroscopic baths with fast internal relaxation times.
As shown in \cite{njp}, Eq. (\ref{eq:lind}) with 
(\ref{eq:hamXY},\ref{eq:bathsc})
can be solved exactly in terms of {\em normal master modes} (NMM) 
which are obtained from diagonalization of $4n\times 4n$ matrix $\mm{A}$
written in terms of $4\times 4$ blocks
\begin{eqnarray}
\mm{A}_{l,m}&=& \delta_{l,m}(-2h\mm{R}_0
+\delta_{l,1}\mm{B}_{\rm L} + \delta_{l,n}\mm{B}_{\rm R}) \label{eq:bigA}\\
&+& \delta_{l+1,m}\mm{R}_\gamma - \delta_{l-1,m}\mm{R}^T_\gamma,\quad
l,m=1,\ldots,n,\nonumber
\end{eqnarray}
where $\mm{R}_\gamma=\one_2 \otimes (\ii \sigma^y - \gamma \sigma^x)/2$ and
$\mm{B}_{\lambda} = -\frac{1}{2}(\Gamma^\lambda_2+\Gamma^\lambda_1)\sigma^y \otimes \one_2
+\frac{1}{2}(\Gamma^\lambda_2-\Gamma^\lambda_1)(\sigma^z +\ii \sigma^x)\otimes \sigma^y$.

Following \cite{njp}, the key concept is $4^n$ dimensional 
{\em Fock space of
operators} ${\cal K}$ spanned by an orthonormal basis
$P_{\alpha_1,\alpha_2,\ldots,\alpha_{2n}} := w_1^{\alpha_1}w_2^{\alpha_2}\cdots w_{2n}^{\alpha_{2n}},\,
\alpha_j\in\{ 0,1\}$ where $w_{2m-1} = \sigma^x_m \prod_{m'<m} \sigma^z_{m'},
w_{2m} = \sigma^y_m \prod_{m'<m} \sigma^z_{m'}$ are $2n$ anticommuting Majorana operators $\{w_j,w_k\} = 2\delta_{j,k}$. We introduce 
{\em canonical adjoint Fermi maps} over ${\cal K}$,
defined as
$
\cc_j \xket{P_{\un{\alpha}}} = \delta_{\alpha_j,1} \xket{w_j P_{\un{\alpha}}}
$,
so the quantum Liouvillean (\ref{eq:lind}) becomes {\em bilinear}
$\LL = \un{\aaa}\cdot\mm{A}\un{\aaa} + {\rm const}\one$
in Hermitian maps $\aaa_{2j-1} = \frac{1}{\sqrt{2}}(\cc_j + \cc^\dagger_j),\aaa_{2j} = \frac{\ii}{\sqrt{2}}(\cc_j - \cc^\dagger_j)$, satisfying $\{\aaa_p,\aaa_q\}=\delta_{p,q}$.
Note that the eigenvalues of $4n\times 4n$ antisymmetric matrix $\mm{A}$ 
(\ref{eq:bigA})
called {\em rapidities} come in pairs $\beta_1,-\beta_1,\beta_2,-\beta_2,\ldots,\beta_{2n},-\beta_{2n}$, $\re\beta_j\ge 0$.
The corresponding eigenvectors $\un{v}_p,p=1,\ldots,4n$, defined by
$\mm{A}\un{v}_{2j-1}=\beta_j \un{v}_{2j-1}$,
$\mm{A}\un{v}_{2j}=-\beta_j \un{v}_{2j}$, can always be normalized as
$\un{v}_{2j-1}\cdot \un{v}_{2j} = 1$ and $\un{v}_p\cdot \un{v}_q=0$ 
otherwise. Writing NMM maps as 
$\bb_j = \un{v}_{2j-1}\cdot \un{\aaa}$, 
$\bb'_j = \un{v}_{2j}\cdot \un{\aaa}$, in general $\bb'_j\neq \bb^\dagger_j$, 
obeying almost-canonical anticommutations 
$\{\bb_j,\bb_k\}=\{\bb'_j,\bb'_k\}=0,\{\bb_j,\bb'_k\}=\delta_{j,k}$,
the Liouvillean (\ref{eq:lind}) takes the {\em normal form}, 
$\LL = -2\sum_{j=1}^{2n} \beta_j \bb'_j \bb_j$.
Thus a complete set of $4^n$ eigenvalues of $\LL$ (real parts being the
relaxation rates) can be constructed as $-2\sum_j \nu_j\beta_j$
where $\nu_j\in\{0,1\}$ are eigenvalues of $2n$ {\em mutually commuting,
non-hermitian} number operators $\bb'_j \bb_j$.

Let $\ness$ be the element of ${\cal K}$ corresponding to the stationary solution $\rho_{\NESS}$ (NESS) of Eq. (\ref{eq:lind}), i.e. zero eigenvalue of $\LL$,
$\nu_j\equiv 0$. 
The main result of \cite{njp} (Th.~3) takes into account the fact
that $\ness$ is a right-vacuum of $\LL$ -- the left-vacuum being the trivial identity-state
$\ket{\one}$ -- and asserts that any quadratic physical observable can be explicitly 
computed in terms of eigenvectors $\un{v}_p$,
$\tr(w_j w_k \rho_{\NESS}) = \delta_{j,k} + \bra{\one}\cc_j \cc_k\ness$,
\begin{eqnarray}
&&\bra{\one}\cc_j \cc_k\ness = \frac{1}{2}\sum_{m=1}^{2n}\!\bigl(
v_{2m,2j-1} v_{2m-1,2k-1}- \label{eq:cor2} \\
\!\!\!\!\!\!&\!\!-\!\!&\!v_{2m,2j}v_{2m-1,2k}\!\!-\!\!\ii v_{2m,2j}v_{2m-1,2k-1}\!\!-\!\!\ii v_{2m,2j-1}v_{2m-1,2k}\!\bigr). 
\nonumber
\end{eqnarray}
Higher order observables can be computed using the Wick theorem. For example, 
noting $\sigma^z_m = -\ii w_{2m-1}w_{2m}$,
{\em spin-spin correlator} which we shall study later reads
\begin{eqnarray}
C_{l,m} &=& \tr(\sigma^z_l \sigma^z_m \rho_{\NESS}) - 
\tr(\sigma^z_l \rho_{\NESS})\tr(\sigma^z_m \rho_{\NESS}) \label{eq:corr}\\
&=&\bra{\one}\cc_{2l-1} \cc_{2m}\ness \bra{\one}\cc_{2l} \cc_{2m-1}\ness\nonumber\\ 
&-&\bra{\one}\cc_{2l-1} \cc_{2m-1}\ness \bra{\one}\cc_{2l} \cc_{2m}\ness\;\;{\rm if}\;l\neq m.
\nonumber
\end{eqnarray} 
As proven in \cite{njp},
NESS is {\em unique} iff rapidity spectrum is non-degenerate, $\beta_j\neq 0$
for all $j$, and (almost) any initial state aproaches NESS asymptotically
exponentially with the rate $\Delta = 2\min_j \re \beta_j$ if $\Delta > 0$.

\begin{figure}
         \centering	
	\includegraphics[width=\columnwidth]{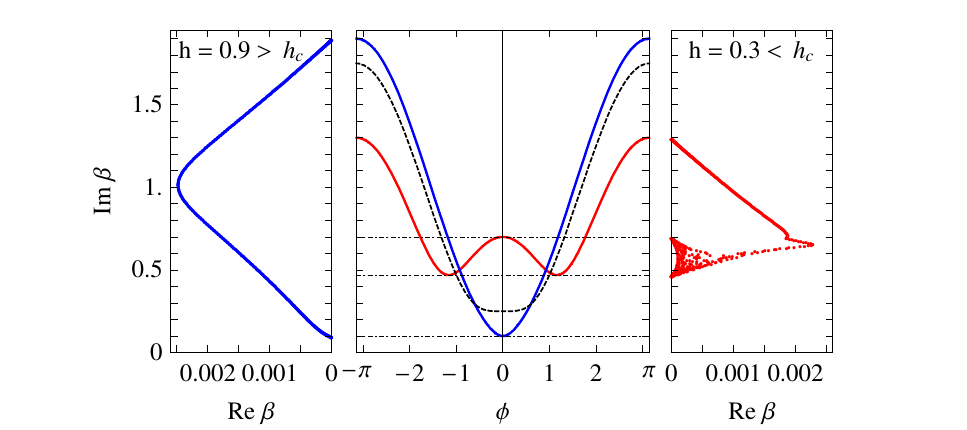}
	\caption{Rapidity spectrum $\{\beta_j\}$ around the imaginary axis for 
$n=640, \Gamma^{\rm L}_1=0.5,\Gamma^{\rm L}_2=0.3,
\Gamma^{\rm R}_1=0.5,\Gamma^{\rm R}_2=0.1, 
\gamma=0.5$, and $h=0.3 < h_c$ (left, blue) and 
$h=0.9 > h_c$ (right, red), compared to dispersion
(\ref{eq:disp}) of the free XY model (center), where dashed 
curve indicates the critical case $h=h_c=0.75$.
	}
	\label{fig:spectrum}
\end{figure}

Let us now proceed to detailed analytical and numerical 
investigation of the structure of NESS in XY chain.
The bulk spectrum of rapidities for $n\to\infty$ is insensitive to the coupling to the
baths and is given by $\beta = \pm \ii \epsilon(\phi)$, $\phi\in (-\pi,\pi]$ where 
\begin{equation}
\epsilon(\phi) = \sqrt{(\cos\phi - h)^2 + \gamma^2 \sin^2\phi}
\label{eq:disp}
\end{equation}
is the quasi-particle dispersion relation in an infinite XY chain 
(see e.g. \cite{mccoy}). 
For a finite chain (\ref{eq:hamXY}) with the bath coupling on the edges 
(\ref{eq:bathsc}) we find that the
bulk (nearly continuous) rapidity spectrum gains a small {\em never vanishing} 
real part $\re \beta(\phi) = {\cal O}(n^{-1})$.
At the spectral edges $\beta^*$, $\beta^*\vert_{n=\infty}=\pm\ii\epsilon(\phi^*)$, with $\phi^*$ 
defined by $\dd\epsilon(\phi^*)/\dd\phi=0$, 
the {\em gap} is actually much smaller $\re \beta^* = {\cal O}(n^{-3})$
(analytical result, generalizing \cite{njp}). Thus, the asymptotic 
relaxation time to NESS $1/\Delta={\cal O}(n^3)$ diverges in the 
thermodynamic limit $n\to\infty$.

\begin{figure}
	\hspace{-4.5cm}
	\includegraphics[width=1.35\columnwidth]{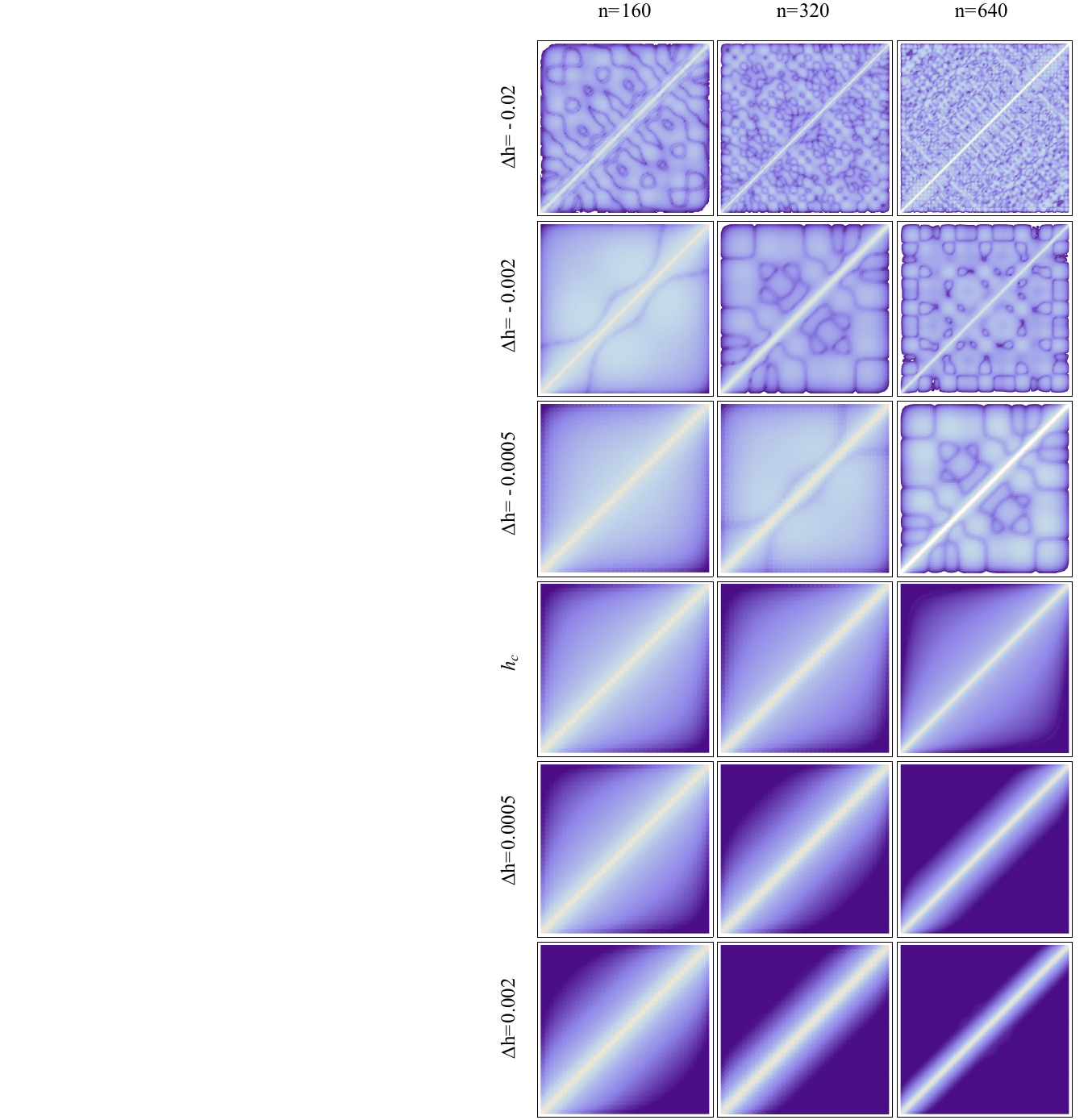}\hspace{1cm}	
\caption{Spin-spin correlation matrices of NESS (\ref{eq:corr}) for three different sizes $n$ (columns)
         and different values of $\Delta h=h-h_c$ (rows) closely surrounding the critical value (\ref{eq:hc}). 
$\gamma,\Gamma^\lambda_\mu$ same as in Fig.~\ref{fig:spectrum}.
 Color-scale is proportional to $\log |C_{l,m}|$ and ranges from 
         $\log 10^{-18}$ (dark-blue) to $\log 1$ (white).}
	\label{fig:cfgrid}
\end{figure}

We note, however, that the structure of the quasi-particle spectrum $\epsilon(\phi)$ qualitatively changes
as the magnetic field crosses a critical value
\begin{equation}
h_c(\gamma) = 1 - \gamma^2,
\label{eq:hc}
\end{equation}
namely for $h < h_c$ the minimal quasi-particle energy exists for a nontrivial value of
quasi-momentum $\phi^* = \arccos[h/h_c(\gamma)]$ 
yielding a new, non-trivial band edge $\beta^*$, whereas for $h > h_c$ the band edges can exist only at points $\phi^*=0,\pi$ (see Fig.~\ref{fig:spectrum}).
Consequently, complex rapidities of an open XY chain shape up a third condensation point near the imaginary 
axis for $h < h_c$ which is composed of NMMs (eigenvectors of $\mm{A}$) with 
pseudo-momenta near $\phi^*\neq 0,\pi$ and has a 
dramatic effect on the structure of NESS as we demonstrate below.

Indeed, as $h < h_c$, we find the emergence of 
{\em long range magnetic correlations} (LRMC) characterized by 
non-decaying structures in the correlation matrix 
$C_{l,m}$ (\ref{eq:corr}). Typical size $\ell$ of the correlation patches
is of the order $\ell \sim 1/\phi^*$ (Fig.~\ref{fig:cfgrid}). For $h\approx h_c$ one finds
critical scaling $\phi^* \approx [2(h_c-h)/h_c]^{1/2}$ which agrees with the data.

 \begin{figure}
         \centering	
	\includegraphics[width=\columnwidth]{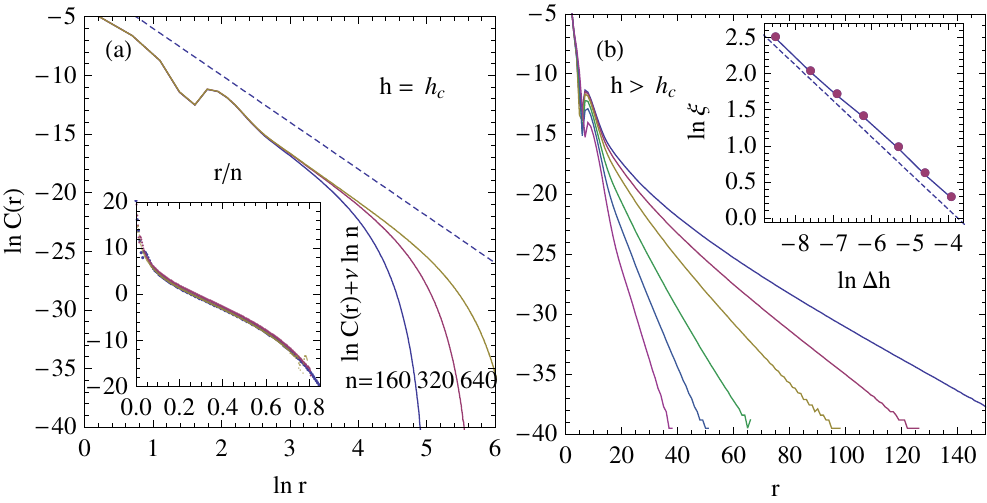}
	\caption{Spin-spin correlator $C(r)$ computed as an 
average of $C_{l,m}$ with fixed $r=|l-m|$ and $|l+m-n| \le 0.08 n$.
In (a) we plot $C(r)$ in the critical case $h=h_c=0.75$ ($\gamma,\Gamma^\lambda_\mu$ same as in Fig.~\ref{fig:spectrum}) for several 
sizes $n=160,320,640$ (indicated), while dashed line indicates asymptotic 
$r^{-4}$ decay (double log-scale).
Inset shows the scaled correlator
$n^\nu C(r)$ versus $r/n$ with $\nu=4.09$ for the same data (normal-log scale).
In (b) we plot $C(r)$ for changing $h = 0.7505, 0.751, 0.752, 0.755, 0.76,
0.77 > h_c$ (right-to-left colored curves) in (log-normal scale) indicating exponential
decay. Inset shows numerically determined localization length $\xi$ 
versus $\Delta h = h-h_c$ (points) as compared to theoretical estimate (\ref{eq:xi}) 
(dashed line).} \label{fig:cfscal}
\end{figure}

In the critical case $h = h_c$ (see Fig.~\ref{fig:cfscal}a) one finds {\em 
power-law decay} of the 
correlation matrix $C_{l,m} \propto |l-m|^{-4}$ if neglecting finite size/boundary effects. 
If we scale the distance we find numerically a finite size scaling 
$n^\nu C_{l,m}=f(|l-m|/n)$ where $\nu=4.09$ and $f(x)$ is some function describing data for 
all large $n$ 
(inset of Fig.~\ref{fig:cfscal}a). Critical point $h=h_c$ is also characterized by 
{\em faster} closing of the spectral gap $\Delta$ of Liouvillean, namely there 
we find $\re\beta^* = {\cal O}(n^{-5})$, meaning $n^2$ times longer 
relaxation times of generic solutions $\rho(t)$ of (\ref{eq:lind}).

For $h > h_c$, we have $\phi^* = 0$ and {\em no} LRMC in NESS. 
Then one finds an {\em exponential decay} of the correlation matrix $C_{l,m} \propto \exp(-|l-m|/\xi)$
with the localization length which can be estimated theoretically from 
a scattering problem defined by the matrix (\ref{eq:bigA}):
\begin{equation}
\xi^{-1} = 4 \cosh^{-1}(h/h_c) \approx 4 [2(h-h_c)/h_c)]^{1/2}
\label{eq:xi}
\end{equation}
where factor $4$ reflects the fact that $C_{l,m}$ is a 4-point function in 
NMM amplitudes $\un{v}_p$ (see Fig.~\ref{fig:cfscal}b).

 \begin{figure}
 	\includegraphics[width=0.8\columnwidth]{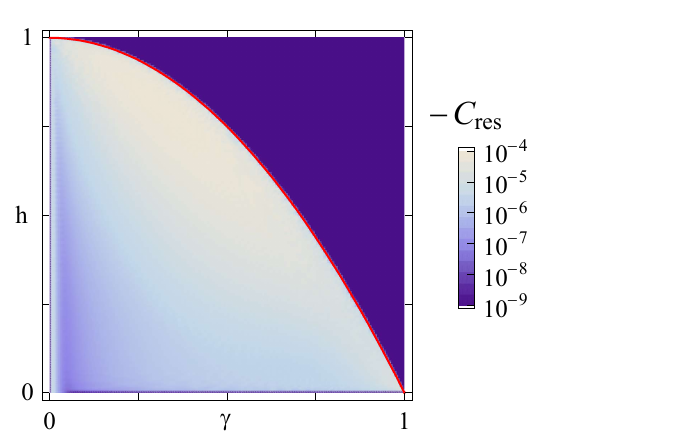}\hspace{-3cm}
	\caption{Phase diagram showing the residual correlator 
$-C_{\rm res}$ (log-scale indicated) 
calculated on $200\times 200$ grid of parameter space 
	$h,\gamma$ for size $n=160$ ($\Gamma^\lambda_\mu$ same as in
Fig.~\ref{fig:spectrum}). Red curve is the 
critical line $h_{\rm c}=1-\gamma^2$. Note that $C_{\rm res}$ is practically
insensitive to increasing the size $n$ in LRMC phase (bright).
 	}
	\label{fig:phase}
\end{figure}

The above results are summarized in a {\em nonequilibrium phase diagram} of XY chain 
(Fig.~\ref{fig:phase}) showing the {\em residual correlator} 
$C_{\rm res} = \sum_{l,m}^{|l-m|>n/2}C_{l,m}/\sum_{l,m}^{|l-m|>n/2}1$ (which is found to be always 
negative) in the $\gamma$-$h$ plane, with the critical curve $h_c(\gamma)$ separating the two phases.
Note that the other boundary lines $\gamma=0$ (XX chain) and $h=0$ (XY with zero field) are {\em not}
in LRMC phase.

In analogy to equilibrium QPTs \cite{osterloh,latorre}, we wish to characterize
the non-equilibrium transition in terms of quantum information theoretic concept, namely 
with the {\em difficulty of classical simulation} of $\rho_{\NESS}$ which is described in terms of OSEE \cite{iztok} (or block-entropy in ${\cal K}$), i.e. von Neumann entropy 
$S(n) = -\tr_{[1,n/2]}\hat{R} \log_2 \hat{R}$ of the reduced density matrix of a {\em half-chain} 
$\hat{R}=\tr_{[n/2+1,n]}\ket{\NESS}\bra{\NESS}$. 
$\tr_{[j,k]}$ corresponds to a partial trace over the sublattice $[j,k]$.
Straightforward calculation, combining Refs. \cite{latorre,njp},
results in
$S(n) = -\sum_{j=1}^n(
(\half +\eta_j)\log_2(\half+\eta_j)+
(\half -\eta_j)\log_2(\half-\eta_j))$,
where $\eta_j$ are $n$ {\em positive} eigenvalues of an upper-left (or lower-right) 
$2n\times 2n$ \cite{foot} block of $4n\times 4n$ Hermitian
correlation matrix $\mm{D}_{p,q}=\bra{\NESS}\aaa_p \aaa_q\ket{\NESS}/\braket{\NESS}{\NESS}$.
$\mm{D}$ can be computed by expressing $\aaa_p$ in terms of NMM maps
$\bb_j$ and $\bb^\dagger_j$ (not $\bb'_j$), $
\un{\aaa} = \mm{Q}^*\un{\bb} + \mm{Q}\un{\bb}^\dagger$.
Namely, $\mm{D} = \mm{Q}^* \mm{T} \mm{Q}^T$, where $\mm{T}_{j,k}=\sum_{p=1}^{4n} v_{2j-1,p} v^*_{2k-1,p}$ is
$2n\times 2n$ matrix, and $\mm{Q}=\mm{V}_{\!o} \mm{K}_{12}$ where
$\mm{K}_{12}$ designates upper-right $2n\times 2n$ quarter of $4n\times 4n$ matrix
$\mm{K} = -(\mm{V}_{\!o}|\!-\!\!\mm{V}_{\!o}^*)^{-1} (\mm{V}_{\!e}|\!-\!\!\mm{V}_{\!e}^*)$ and 
$(\mm{V}_{\!e})_{p,k} = v_{2k,p},(\mm{V}_{\!o})_{p,k}=v_{2k-1,p}$ are $4n\times 2n$ matrices.
$(\mm{X}|\mm{Y})$ denotes vertical concatenation of two $4n\times 2n$ matrices
into a single $4n\times 4n$ matrix. 

\begin{figure}
         \centering	
	\includegraphics[width=\columnwidth]{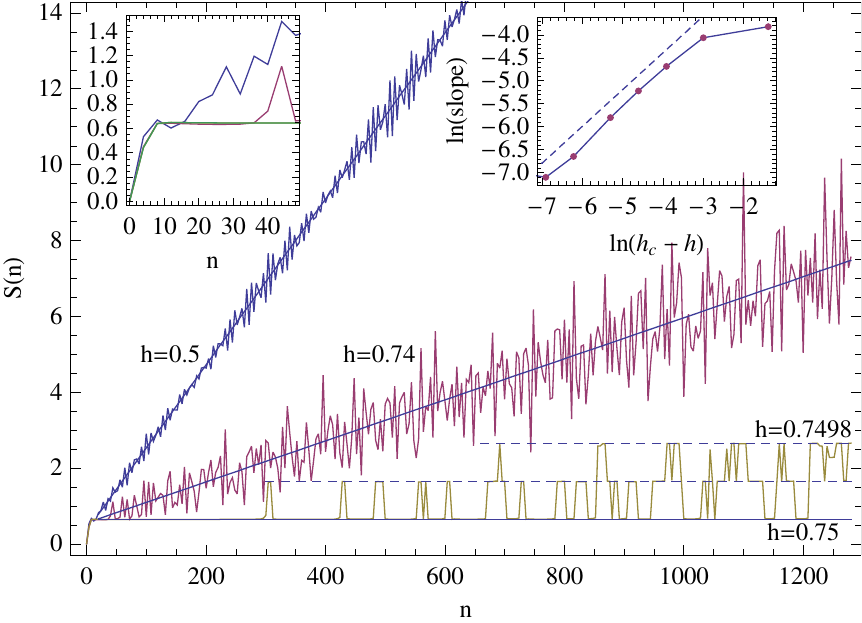}
	\caption{
OSEE $S(n)$ (symmetric chain bipartition) for $\gamma,\Gamma^\lambda_\mu$ of
Fig.~\ref{fig:spectrum} and different $h \le h_c=0.75$ 
(indicated). 
Best fitting linear growths are indicated with straight lines. Dashed horizontal lines indicate 
$S_0 + 1$ and $S_0+2$, $S_0$ being the saturation value for $h=h_c$.
Left inset just magnifies the scale while right inset shows the slope of $S(n)$ growth vs.
$h_c-h$ (log-log) and dashed line indicates $|h_c-h|^{0.8}$.
}
\label{fig:oseeslopes}
\end{figure}

The resulting behaviour of $S(n)$ in NESS of XY chain is striking 
(see Fig.~\ref{fig:oseeslopes}): LRMC phase $h < h_c$ is characterized with
a linear growth $S(n) = s n + {\rm const}$, with some constant $s>0$. This has to be contrasted with a $\log n$ growth found for equilibrium critical models \cite{latorre}. 
As $h$ approaches $h_c$ the slope $s$ approaches $0$ as 
$s \propto (h_c-h)^\tau$, with numerically determined critical exponent
$\tau \approx 0.80$, and the fluctuations of $S(n)$ around an average linear growth increase. 
These fluctuations can be explained by sensitive dependence of NESS on boundary 
conditions (bath couplings or size changes) due to long range correlations, evident also in the structures
of the correlation matrices (Fig.~\ref{fig:cfgrid}). Note also an interesting 'quantization of bipartite entanglement' which is observed
for very small $h_c-h$ where $S(n)$ can take only approximately a discrete set of values 
$S(n) \approx S_0 + k, k\in \ZZ^+$ and which can be explained by the
quasi-particle 
picture of NMM. At and above the critical field $h \ge h_c$ we find {\em saturation} 
$S(n) = {\cal O}(1)$, and vanishing fluctuations of $S(n)$ since there NESS becomes insensitive to
boundary conditions due to fast decay of magnetic correlations. {\em Only} there can NESS be efficiently simulated, 
e.g. in terms of matrix product states \cite{schuch}, by numerical methods
like {\em density matrix renormalization group} (DMRG) \cite{dmrg}. 

All the numerical results presented above have been obtained for a fixed non-equilibrium bath
couplings $\Gamma^{\rm L}_1=0.5,\Gamma^{\rm L}_2=0.3,\Gamma^{\rm R}_1=0.5,\Gamma^{\rm R}_2=0.1$. However, the
results {\em did not change} qualitatively, in particular the phase boundary, 
when we (i) varied the bath couplings $\Gamma^\lambda_\mu$ \cite{wichterich}, 
(ii) coupled {\em several spins}
around each end to Lindbladian baths, or (iii) even set the bath couplings equal 
$\Gamma^{\rm L}_\mu = \Gamma^{\rm R}_\mu$. The latter case (iii) does not represent 
equilibrium situation, i.e. $\rho_{\NESS}$ is not a {\em thermal state} 
$\rho_{T}=Z^{-1}\exp(-H/T)$ as the XY chain is not ergodic \cite{mccoy}.
For example, no discontinuity at $h=h_c$ appears in the properties of $\rho_T$
for any $T$, and correlator $C(r)$ essentially always decays with 
$T-$dependent rates \cite{mccoy}, whereas in non-LRMC phase of 
NESS decay length $\xi$ is asymptotically insensitive to bath 
parameters (\ref{eq:xi}). Furthermore, thermal states in one-dimension have 
always bounded (in $n$) OSEE \cite{PPZ}, and related quantities like 
{\em mutual information} \cite{cirac},
hence the simulation complexity of NESS is qualitatively different.

In spite of demonstrated discontinuity in the spin-spin correlation function, the local 
observables such as energy or spin 
density in NESS are numerically found to be smooth functions of $h$ at $h_c$, so
the non-equilibrium transition appears to be of high or infinite order 
(similar to Kosterlitz-Thouless transition). LRMC phase could perhaps be difficult to detect experimentally 
as the residual correlation 
$C_{\rm res}$ is not larger
than few times $10^{-4}$ (Fig.~\ref{fig:phase}) even in the optimal case 
(w.r.t. varying $\Gamma^\lambda_\mu$). 

In conclusion, we report on the QPT in NESS of open 
quantum XY spin chain, whose theoretical and numerical description is 
formally analogous to 
equilibrium QPTs in spin chains at zero temperature inasmuch as
NESS can formally be treated as a `ground state' of the quantum Liouvillean.
We show that the phase transition is of mean-field type as the quasi-particle picture gives
a satisfactory theoretical description, in particular the phase boundary between
long-range and exponentially decaying magnetic correlations. 
We have demonstrated that the
two phases, respectively, correspond to linearly growing and saturating entanglement entropy
of NESS in operator space as a function of the chain length. This behavior is
drastically different than in equilibrium quantum XY chains. 
We thank M. \v Znidari\v c for useful comments and independent verifications of the results on
small systems with DMRG \cite{dmrg} codes. 
The work is supported by grants P1-0044 and J1-7347 of Slovenian Research Agency.


\begin{thebibliography}{10}
\bibitem{foot}
OSEE $S$ for a smaller block of $k$ spins (other than a half chain) is simply obtained by taking
the corresponding $4k\times 4k$ part of matrix $\mm{D}$.
\end{thebibliography}
\end{document}